\documentclass[journal,10pt]{IEEEtran}
\usepackage{amsfonts}
\IEEEoverridecommandlockouts

\ifCLASSINFOpdf
\else
\fi

\usepackage{amssymb}
 \usepackage{amsthm}
\usepackage{epsfig}
\usepackage{graphicx}
\usepackage{psfig}
\usepackage{epsf}
\usepackage[cmex10]{amsmath}
\usepackage{booktabs}
\usepackage{fancyhdr}
\usepackage{booktabs}
\usepackage{stfloats}
\usepackage{color}
\usepackage[ruled,linesnumbered]{algorithm2e}
\usepackage{url}
\usepackage{xcolor}
\usepackage{amsmath}

\begin{document}
\title{A Novel Visible Light Communication System Based on a SiPM Receiver}
\author{\IEEEauthorblockN{Zhenzhou Deng, Anyi Li}
\thanks{
Zhenzhou Deng is from School of Life Science and technology, Huazhong University of Science and Technology, Wuhan, China.}
\thanks{
Anyi Li, Zhitao Liu are from School of Qianhu, Nanchang University, Nanchang, China.}

}
\maketitle
\begin{abstract}
Si-Photomultiplier (SiPM) has several attracting features, which can be helpful in the communication field, such as high photon detection efficiency, fast transient response, excellent timing resolution, and wide spectral range. In this paper, we compare SiPM with photodiode (PD), Avalanche photodiodes (APD), and Photomultiplier tube (PMT) in terms of the transformation related performance,  and then implement Si-Photomultiplier (SiPM) based visible light communication (VLC) system driven with visible light light-emitting diode (LED). The system throughout ability, transmission rate, data reconstruction, and required AD sample rate are evaluated. The encouraging results suggest that the SiPM receiver has great application potentials, such as optical wireless communication systems and light fidelity, in which a wide bandwidth of the sensor response is important to enhance the transfer rate.
\end{abstract}
\vbox{} 
\IEEEpeerreviewmaketitle
\section{Introduction}
Self-sustainable green-smart houses \cite{balta2013development}, optical wireless communication (OWC) \cite{elgala2011indoor} and light fidelity (Li-Fi) \cite{tsonev2014light} have attracted considerable attention, in which users always demand large data capacity and multi-functional lighting control during daily life. In contrast to traditional Wi-Fi and fiber-optic communications, the visible light communication (VLC) system \cite{komine2004fundamental} is a complementary system with advantages of customizable space, license free, electromagnetic immunity, communication safety and so on. The VLC has found its suitable application scenarios in mobile connection \cite{ling2004mobile}, indoor positioning \cite{liu2007survey}, vehicle transportation \cite{national1997toward}, targeted communication \cite{manchanda2008role}, underwater resource exploration \cite{akyildiz2005underwater} and hospital/healthcare applications \cite{aminian2013hospital,pathak2015visible,biton2018improved,islim2017towards,haas2015visible}.

At this stage, some of the conceptual VLC prototypes have already been implemented in practice for commercial applications. Several worldwide companies including Bytelight, Target, Emart, and Royal Philips NV have successfully guided shoppers to goods based on their position in retail stores, in which light-emitting diodes (LEDs) \cite{schubert2018light} based VLC are employed to communicate with the camera of smartphones. Other than these applications, intelligent medical care system \cite{cahyadi2015patient} using VLC has been established and working in several hospitals through the cooperation with the Industrial Technology Research Institute (ITRI), which stably and precisely realizes the wireless streaming of therapeutic data and the positioning sensing of medical personnel. According to the report from Grand View Research, all the expected revenue of the global VLC market will reach up to USD 101.30 billion by 2024 \cite{chi2017violet, wang2017study, khan2017visible, chi2015450, janjua2015going}.

\begin{figure}[!ht]
\center
\includegraphics[width=3.5in]{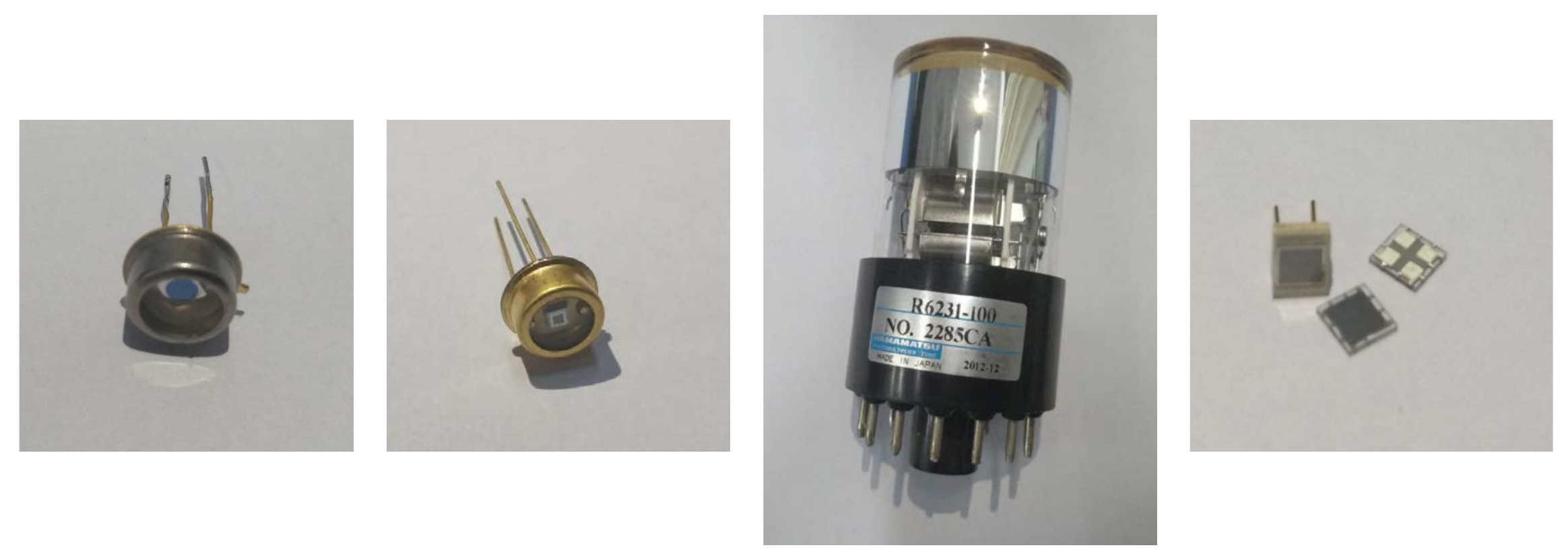}%
\caption{Four different photon sensors: PD, APD, PMT, SiPM, from left to right. We can see the PMT has much more bulky size than other photo sensors. SiPM use surface mount package which will be feasible in more applications.}
\label{Senser}
\end{figure}

\begin{figure}[!ht]
\center
\includegraphics[width=3 in]{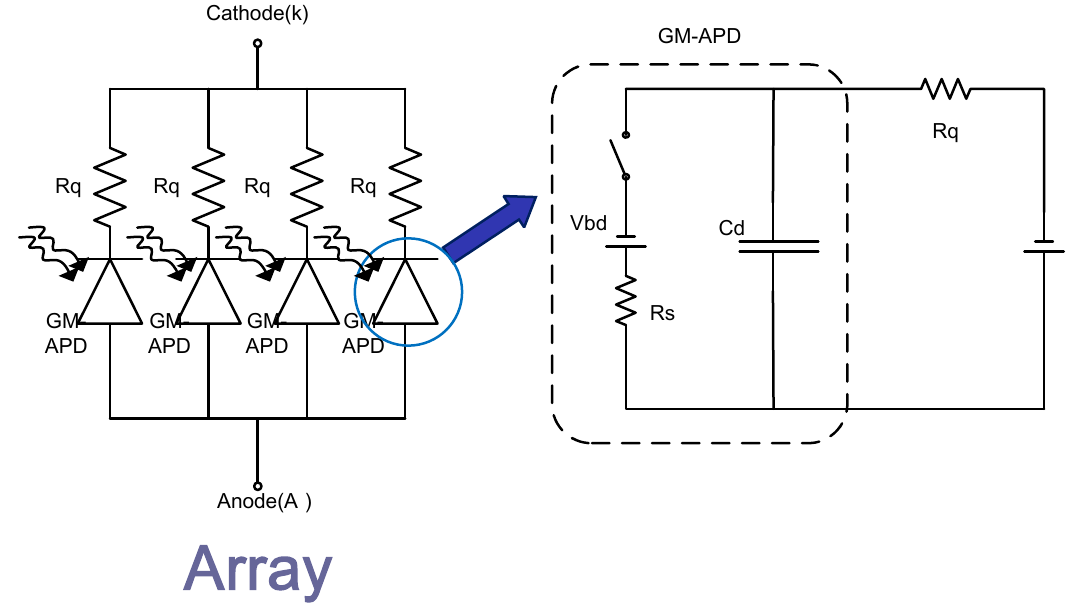}%
\caption{The inside structure of SiPM. Schematics of a modern silicon photomultiplier (SiPM) is composed of an array of single photon avalanche diode microcells with passive quenching. It makes SiPM output recovers in short time, and has a greater slew rate on the pulse leading edge.}
\label{SiPMstruct}
\end{figure}

Almost all these VLC applications employ photodiode (PD) to achieve photo-electronic transition and receive data. However, PD still has many shortcomings, such as small area, no internal gain, much lower overall sensitivity (the use of photon counting is limited, usually used for cooling with special electronic circuit photodiodes). Silicon photomultipliers (SiPMs) \cite{dinu2007development,corsi2007modelling}, also identified as multi-pixel photon counters shown in Fig. \ref{SiPMstruct}, are a favorable class of semiconductor-based photodetectors addressing the challenge of detecting, timing and quantifying low-light optical signals down to the single-photon counting level. SiPMs offer a highly attractive alternative that extremely increases the detection sensitivity of photo-electronic transition while providing all the benefits of PD \cite{eraerds2007sipm,braga2014fully,dinu2007development,powolny2011time,varkonyi2016data}.

In this paper, we will focus on a new particular application of SiPM to the visible light communication field. In the next sections, we will discuss the choice cause of SiPM for VLC among different sensors in Fig. \ref{Senser}. Next, we develop a SiPM based VLC prototype link. Then the related performances were investigated in Section V. At the end, we provide a summary for the VLC application of SiPM.

The following  abbreviations used in this manuscript are list in Table. \ref{Abbreviation}.

\begin{table}[!t]
\caption{Feature Comparison with Other Detector Technologies}
\label{TFeature}
\centering
\begin{tabular}{c c c c c}
\hline
Features &  PD/PIN & APD & PMT & SiPM\\
\hline
Gain & \color{red}{$1$}  & $10^2$   &\color{green}{$10^6$}  & \color{green}{$10^6$}\\
Operational Bias &\color{green}{Low}&\color{red}{High}&\color{red}{High}&\color{green}{Low}\\
Temp.Sensitivity&\color{green}{Low}&\color{red}{High}&\color{green}{Low}&\color{green}{Low}\\
Package Size&\color{green}{Compact}&\color{green}{Compact}&\color{red}{Bulky}&\color{green}{Compact}\\
Mechanical Robustness&\color{green}{Heigh}&\color{red}{Medium}&\color{green}{Low}&\color{green}{High}\\
Ambient light exposure?&OK  &OK&\color{red}{NO}&OK\\
Spectral range&Red  &Red&Blue/UV  &Blue\\
Readout/ Electronics&\color{red}{Complex}&\color{red}{Complex}&\color{green}{Simple} &\color{green}{Simple}\\
Large area available? &\color{red}{No}&\color{red}{No}&\color{green}{Yes}&\color{green}{Yes}\\
Sensitive to magnetism?&\color{green}{No}&\color{green}{No}&\color{red}{Yes}&\color{green}{No}\\
Noise&\color{green}{Low}&Medium  &\color{green}{Low}&\color{red}{High}\\
\hline
\end{tabular}
\end{table}

\section{Comparison With Other Detector Technologies}
Silicon photomultipliers (SiPMs, also called SPM, GAPD, MPPC), are a favorable class of semiconductor-based photodetectors addressing the challenge of detecting, timing and quantifying low-light optical signals down to the single-photon counting level. SiPMs offer a highly attractive alternative that keeps the low-light detection capabilities of traditional photomultiplier tubes while providing all the benefits of other solid-state devices.\cite{yepes2017development,castaneda2016high,konstantinou2016experimental,buzhan2006large,bencardino2009development}
During the last decade, the SiPM has become more and more popular as photon sensor, especially for applications where the need for a large sensitive area is not an issue but the photon counting capability, pulse bandwidth, and time resolution are important. VLC is only one of the potential applications, in which high pulse bandwidth guarantees the high transfer rate in the light information system.

Such a photodetector probabilily paves the way to compact photonic systems which can be easily embedded into smaller and smaller frameworks. Innumerable applications are underway or foreseen which make use of SiPM, as for instance in nuclear physics, nuclear medicine, biochemistry, 3D imaging, automotive and several other fields. 

Low-light photon detectors constitute the technology of many rapidly growing, such as radiation detection, nuclear medicine imaging, spectroscopy, fluorescence analysis, quality control, etc. all require detectors for quantifying and/or time-marking optical signals with any place from 1 to 20,000 photons per signal pulse.  The ideal detector provides a response proportional to the incident photon flow and uses an internal gain mechanism to generate a signal large enough to be easily processed. It has sub-nanosecond response time and broad spectral sensitivity, be rugged, easy to operate, and produce only manageable noise or dark counting rates.

Up to now,  the Photomultiplier tube (PMT), a well-established and widely available vacuum tube device, has become the detector of choice for such applications. The semi-transparent photocathode deposited inside the entrance window inherently limits the Photon Detection Efficiency (PDE) that they can achieve, with typical PMTs having about 20\% at 420nm. A gain of $(1-10) \times 10^6$ is achieved at the cost of a high bias voltage of 1-2kV, which requires the use of costly high-voltage power supplies. PMT is generally stable and low noise, but it is bulky and small due to its vacuum tube structure. Magnetic films can also adversely affect them, limiting their suitability for certain applications. The fragile and bulky feature of PMT extremely limited the application of VLC.

Semiconductor devices have many practical advantages over the PMT, and this led to the Positive-Intrinsic-Negative(PIN) diode being used in applications where PMTs are too bulky or delicate, or where high voltages are not possible. However, PIN diodes are severely limited by their complete lack of internal gain. Avalanche photodiodes (APDs) are a relatively new technology for the extension of simple PIN diodes. Here the reverse bias increases to a point where impact ionization allows for some internal multiplication but is below the breakdown bias where the Geiger mode would take over. Thus, at a bias of 100-200V, a gain of about 100 can be obtained. With special manufacture, it is possible for gains of several thousand to be reached using a High voltage bias of more than 1500V. Whilst the gain may be lower than that of a PMT, APDs have the advantage of a PDE which can be $>$65\% and also a compact size, ruggedness, and insensitivity to magnetic fields. Their main drawbacks are their excess noise (associated with the stochastic APD multiplication process) and in an important trade-off: The capacitance increases with the device area and decreasing thickness, whereas the transit times of the charge carriers increase with the thickness, implying a performance trade-off between noise and timing. Their size is limited to 10 mm diameter.

The SiPM has high gain and enough PDE (more than 20\%), very similar to the PMT, but has the physical benefits of compactness, ruggedness and magnetic insensitivity in common with the PD and APD. In addition, the SiPM achieves its high gain ($1 \times 10^6$) with very low bias voltages (~30V) and the noise is almost entirely at the single photon level. Because the high degree of uniformity between the microcells the SiPM is capable of discriminating the precise number of photoelectrons detected as distinct, discrete levels at the output node. The ability to measure a well-resolved photoelectron spectrum is a feature of the SiPM which is generally not possible with PMTs owing to the variability in the gain or excess noise.

To summarize, among all the photosensors in Table. \ref{TFeature}, the extremely remarkable performance achieved by SiPM sensors in terms of high photon detection efficiency, fast transient response, excellent timing resolution, and wide spectral range, has made considerable research activities and technological development to be constantly devoted to SiPM devices within the scientific community involved in medical imaging, high-energy physics, and astrophysics. Here, we aim at the realization of specifically designed LED driven VLC architecture to encode, transfer and decode the light or electrical signals containing information and data. 
\* Because the external electronics need to be close to the detector. \*\* SiPM with SensL, having an operational bias of 30V, meet the requirements of the Extra Low Voltage directive.

\begin{figure*}[!ht]
\center
\includegraphics[width=3.5in]{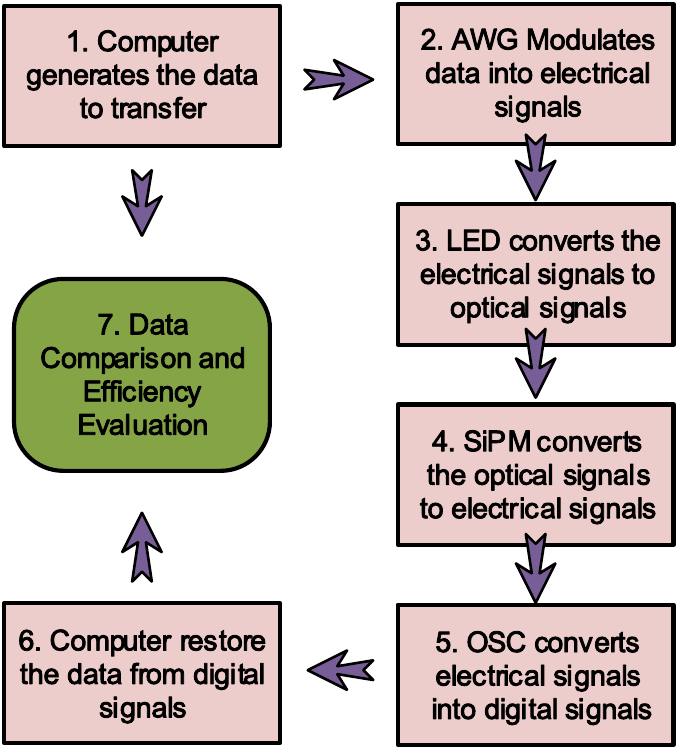}%
\caption{Operating procedure. In the experiment, there were 7 stages: Computer generates the data to transfer,  arbitrary waveform generator(AWG) Modulates data into electrical signals, LED converts the electrical signals to optical signals, SiPM converts the optical signals to electrical signals, Oscilloscope(OSC) converts electrical signals into digital signals ,Computer restore the data from digital signals, Data Comparison and Efficiency Evaluation. }
\label{chartflows}
\end{figure*}

\begin{figure*}[!ht]
\center
\includegraphics[width=5.0in]{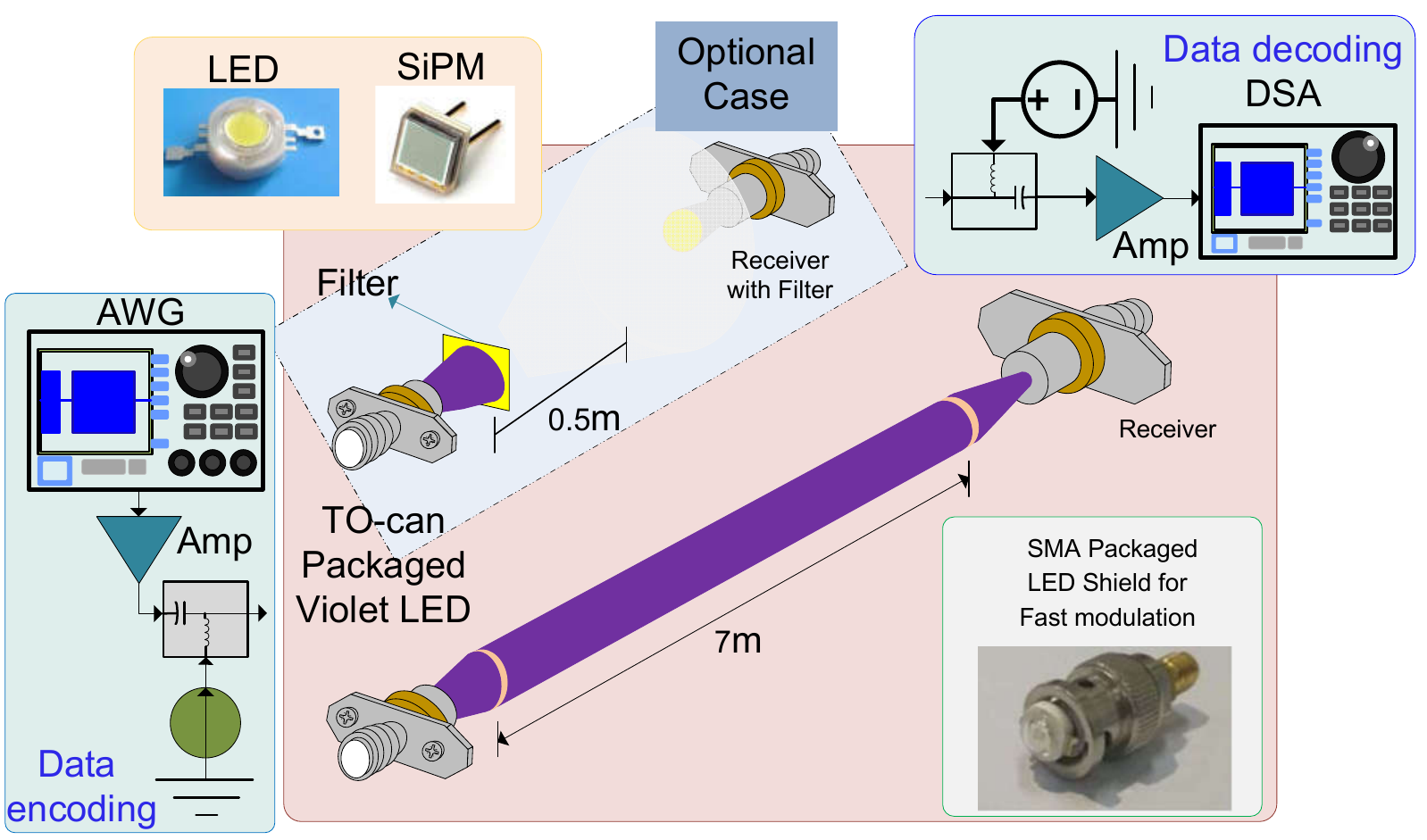}%
\caption{The proposed and implemented VLC system using SiPM. An source LED chip provided by Nanchang University was driven by a current amplifier feed by keysight AWG M8195S.And then an optical fiber optically connected the LED to a MicroFB-30035 SMT Sensl SiPM, while the SiPM signal were output to current amplifier. }
\label{Transfer}
\end{figure*}

\begin{figure*}[!ht]
\center
\includegraphics[width=5in]{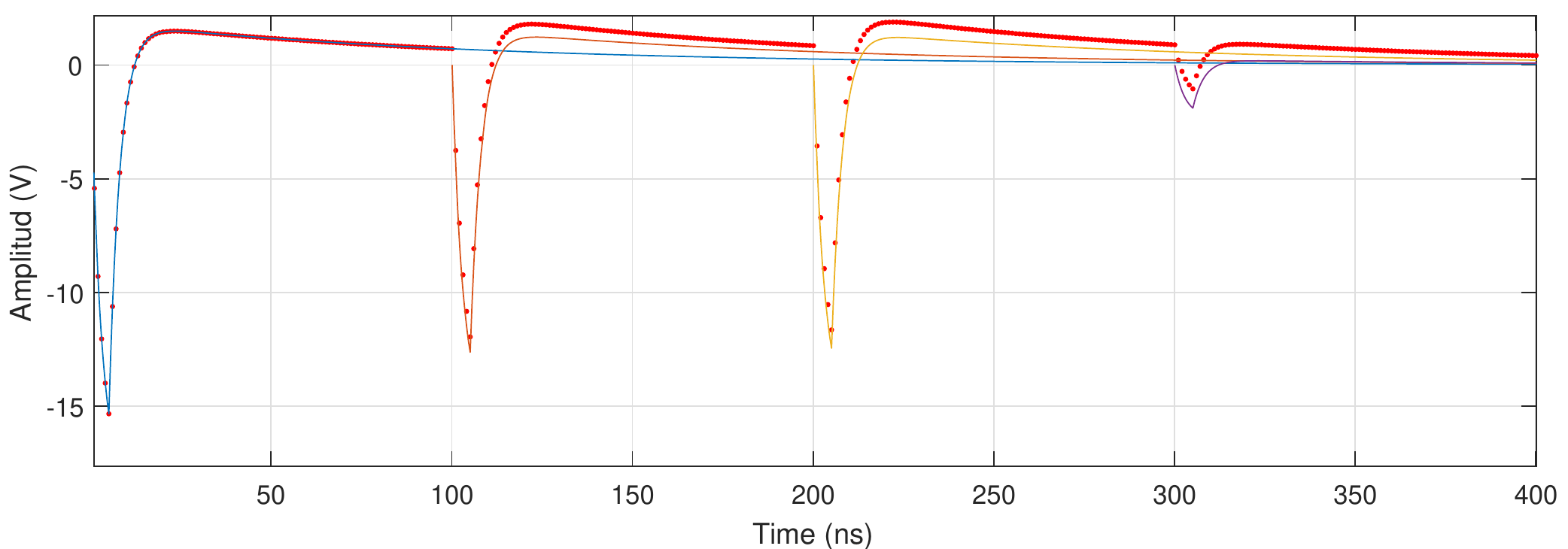}%
\caption{Pulses pile up are common when the repeat frequency are higher than the reciprocal value of the pulse duration. SiPMs always have capacitive characteristic, then produce pulse tails. The SiPM was support with an SMA package connecter and its output was directly connected to a keysight digital storage oscilloscope DSOX6002A with a 50 Ohm termination.}
\label{plotRecon}
\end{figure*}

\section{SiPM based VLC prototype Implementation and Experiment}
Our research team has experience in the design and construction of dedicated positron emission tomography (PET) systems using SiPM, starting with the small animal PET called Trans PET \cite{deng2013empirical,xie2009potentials,zhenzhou2015quadratic,deng2015maximum}, and then the human PET. In order to improve the performance of the photodetector, a sipms array capable of providing high intrinsic spatial resolution and good timing response is designed, and at the same time, a magnetic film can be used. To demonstrate the advance of SiPM detection in VLC, we implement the prototype utilizing SiPM and setup the following experiments.

\subsection{Prototype VLC System Using SiPM}
In the experiment, there were 7 stages as showed in Fig. \ref{chartflows}. Computer generates the data to transfer, arbitrary waveform generator(AWG) Modulates data into electrical signals. LED converts the electrical signals to optical signals. SiPM converts the optical signals to electrical signals. Oscilloscope(OSC) converts electrical signals into digital signals. Computer restores the data from digital signals, Data Comparison, and Efficiency Evaluation.

As shown in Fig. \ref{Transfer}, a source LED chip provided by Nanchang University was driven by a current amplifier feed by keysight arbitrary waveform generator(AWG) M8195S, which can provide 65 GSamples/s and a 20GHz bandwidth, and each 1-slot Axie module has 1, 2 or 4 channel configurations. And then an optical fiber, optically connected the LED to a MicroFB-30035 SMT Sensl SiPM, while the SiPM signal is output to the current amplifier. Despite the fact that the SiPM is sensitive to single photons, its dark count rate of $~100 kHz/mm^2$ at room temperature renders it unsuitable for use for applications at very low light levels. However, with the application of cooling technology (such as with the SensL MiniSL), a two-order reduction in dark count rate is easily achieved. Sensl has more efficient products now, which have more PDE and faster output. We can expect more application when the employing SiPM has better performance. The SiPM is supported with an SMA package connector and its output is directly connected to a keysight digital storage oscilloscope DSOX6002A with a 50 Ohm termination. The oscilloscope is operated with a 6 GHz bandwidth and a 10 GSps sampling rate per channel. At the LED-end, the data are wrapped as electrical pulses serious at different repeat rates: 2 4 5 8 10 20 40 50 80 100 200 400 500MSps. The input data from AWG are encoded as different pulse amplitudes. At the SiPM-end, SiPM pulses containing data are digitized by the oscilloscope. Due to the evidential and fast rise of the pulse shape of SiPM, the sent-in bits can be easy discriminated after digitized. Since all pulses have tails lasting more than 100 ns, a pulse reconstruction processing is required to obtain the transfer data, which can be verified with the recorded original AWG data.
\subsection{Pulse Reconstruction}
Caused by the capacitive characteristic, the transferring pulses always pileup with several adjacent ones as shown in Fig. \ref{plotRecon}. Thus, amplitude of each pulse could be affected when the pileup effect is not considered. To guarantee the received the data consistency and integrality, the pulse reconstruction method is employed.
An amplitude modulated SiPM pulse can be modeled as an impulse function $f_i(t)=
E_{i}\delta(t-t_{i})$, where $\delta(t-t_i)$ is a shifted Dirac Delta function. For the pulse with index $i$, its amplitude
and arrival time are denoted as $E_i$ and $t_i= i \Delta T$, respectively. Here, $\Delta T = 1/ f_r$, where $f_r$ is repeating rate, and $\Delta T$ is pulse interval.  The overall inputs of the system are a series of photons pulse and can be represented as $f(t)=\sum f_i(t)$. The output signals are waveforms denoted as $p(t)$. The VLC system is then expressed as a linear convolution equation
\begin{equation}
\label{pfn}
p(t) = f(t)\ast\varphi(t)+n(t),
\end{equation}
where $\varphi(t)$ is the system's unit impulse response function, $n(t)$ is the noise and $\ast$ is the convolution
operator. If the pulse interval $\Delta t$ of adjacent pulses is too close, there will be pileups in $p(t)$. Otherwise, $p(t)$
contains only single events. To retrieve the transferred data $E_i$, we need to solve the inverse problem of (\ref{pfn}),
and use the collected signal $p(t)$ to compute $f(t)$.

In practice, signals mentioned above are handled in their discrete forms. We denote $\vec{f}=\{f_j\}$,
$\vec{p}=\{p_l\}$ and $\vec{n}= \{n_l\}$ as the discrete forms of the input, output
and noise signal sequences, respectively. The system impulse response function is also expressed as a vector
$\vec{\varphi}=\{\varphi_r\}$, where $\varphi_r=\varphi(r\cdot \Delta t)$ and $\Delta t$ is the sampling interval. $r$,
$j$ and $l$ are element indexes in the corresponding vectors. For the convenience of computation, the convolution
operation in (\ref{pfn}) is rewritten as matrix multiplication. A Toeplitz matrix $H =
T\{\vec{\varphi}\}$ is generated from $\vec{\varphi}$ and the convolution (\ref{pfn}) is transferred to

\begin{equation}
\label{pHfn}
\vec{p} = H\vec{f} + \vec{n}.
\end{equation}
The element of $H$ is denoted as
\begin{equation}
\label{pHfn1}
h_{lj}=\varphi_{l-j},
\end{equation}
where $\varphi_{l-j}$ is the $(l-j)$th element in vector $\vec{\varphi}$. Obtaining $\vec{f}$ from (\ref{pHfn}) is
an inverse problem which can be solved by many kinds of well established
methods, including our proposed Successive Remove (SR) method, pulse model based iterative deconvolution (PMID) method and Fourier deconvolution (FD) method.

In the numerical experiment, the reference method is pulse model based iterative deconvolution \cite{deng2013scintillation} and Fourier deconvolution\cite{brown1992introduction,Nagaoka2018Basic}. PMID method employs MLEM iteration to obtain the pulse amplitude. The general form of MLEM algorithm that expresses the
updating procedure at the $k$th iteration is
\begin{equation}
\label{MLEM1}
f_{j}^{k}= \frac{f_{j}^{k-1}}{\underset{l}{\sum} h_{lj}} \underset{l}{\sum}h_{lj}\frac{p_{l}}{\underset{j}{\sum} h_{lj}
f_{j}^{k-1}},
\end{equation}
where $f_{j}^{k}$ denotes $j$th element of solution $\vec{f}$ at the $k$th iteration. With enough rounds of iterations, the
solution will approach to the original impulse function $\vec{f}$. PMID method can be described as Algorithm \ref{pseudocodePMID}:

\begin{algorithm}[!ht]
\caption{Pulse Model based Iterative Deconvolution (PMID) Algorithm.}\label{pseudocodePMID}
\KwData{Data Sample $\{ S(t),t = t + k \bigtriangleup t$ \}, $\bigtriangleup t$ is the sampling period, $\Delta T$ is pulse interval, $t < N \Delta T$}
\SetKwRepeat{doWhile}{do}{while}
    The start solution $f^{0} = \mathbf{1}$, which is a all-1 vector.\\
    Calculate $\underset{l}{\sum} h_{lj}$.\\
\doWhile{
    $r < M $, where $M$ is the loop times.
}{Forward Projection $\underset{j}{\sum} h_{lj} f_{j}^{k-1}$;\\
  Calculate  $\frac{p_{l}}{\underset{j}{\sum} h_{lj} f_{j}^{k-1}}$;\\
  Backward Projection $ \underset{l}{\sum} h_{lj}\frac{p_{l}}{\underset{j}{\sum} h_{lj} f_{j}^{k-1}}$;\\
  Calculate $\frac{f_{j}^{k-1}}{\underset{l}{\sum} h_{lj}}$;\\
  Calculate $f_{j}^{k}= \frac{f_{j}^{k-1}}{\underset{l}{\sum} h_{lj}} \underset{l}{\sum} h_{lj}\frac{p_{l}}{\underset{j}{\sum} h_{lj}f_{j}^{k-1}}$.
}
Output the $E_k, k = 1,2,3...,N.$ from $f^{M}$.
\end{algorithm}

And Fourier Deconvolution employs regulated Fourier domain division to calculate Wiener Filter as Eq. \ref{Fourier deconvolution}, which executes an optimal tradeoff between inverse filtering and noise smoothing. It eliminates additive noise while inverting blur. The Wiener filtering is optimal in terms of the Mean Square Error(MSE). FD method can be described as Algorithm 2 as: \ref{pseudocodeFourier}:
\begin{equation}
\label{Fourier deconvolution}
f_{FD} = \frac{\digamma{\{p\}} (|\digamma{\{\varphi\}}|^2)}{\digamma{\{\varphi\}} (|\digamma{\{\varphi\}}|^2 +K)}.
\end{equation}
where, $\digamma \{\cdot\}$ denotes Fourier Transform, and $K$ is a constant, which is defined by the noise property.
\begin{algorithm}[!ht]
\caption{Fourier Deconvolution (FD) Algorithm.}\label{pseudocodeFourier}
\KwData{Data Sample $\{ S(t),t = t + k \bigtriangleup t$ \}, $\bigtriangleup t$ is the sampling period}
Calculate the Fourier Transform of $\varphi$;\\
Calculate the Fourier Transform of $p$;\\
Zero padding $\digamma{\{\varphi\}}$ to the same length as $\digamma{\{p\}}$;\\
Calculate $f_{FD} = \frac{\digamma{\{p\}} (|\digamma{\{\varphi\}}|^2)}{\digamma{\{\varphi\}} (|\digamma{\{\varphi\}}|^2 +K)}$;\\
Output the $E_k, k = 1,2,3...,N.$ from $f_{FD}$.
\end{algorithm}

The VLC system employs an Successive Remove (SR) method. Successive Remove (SR) starts with a pulse with ZERO height forward pulses $\{S\{ t \} ,0\leq t < T\}$. Each pulse height estimation follows the Remove operation, which reduces the effect of forwards pulses. SR method can be described as Algorithm \ref{pseudocode}:

\begin{algorithm}[!ht]
\caption{Successive Remove (SR) Algorithm.}\label{pseudocode}
\KwData{Data Sample $\{ S(t),t = t + k \bigtriangleup t$ \}, $\bigtriangleup t$ is the sampling period, $\Delta T$ is pulse interval}
\SetKwRepeat{doWhile}{do}{while}
    The 1st SiPM pulse has no precomponent, then  $E_{1} = \int_0^{\bigtriangleup T} S(t) dt $.\\
\doWhile{
    $t < N \bigtriangleup T $
}{Estimation the precomponent $S_k$ from previous pulses to $k$th pulse;\\
  Calculate the integral $I(k)$ between $(k-1)\bigtriangleup T$ and $k \bigtriangleup T$;\\
  Calculate $E_k$ by removing the precomponent $S(k)$, $E_k = I(k) - S(k)$
}

Output the $E_k, k = 1,2,3...,N.$
\end{algorithm}

\section{Simulation Results}
\subsection{Single Photon Level Detection}
\begin{figure}[!ht]
\center
\includegraphics[width=3 in]{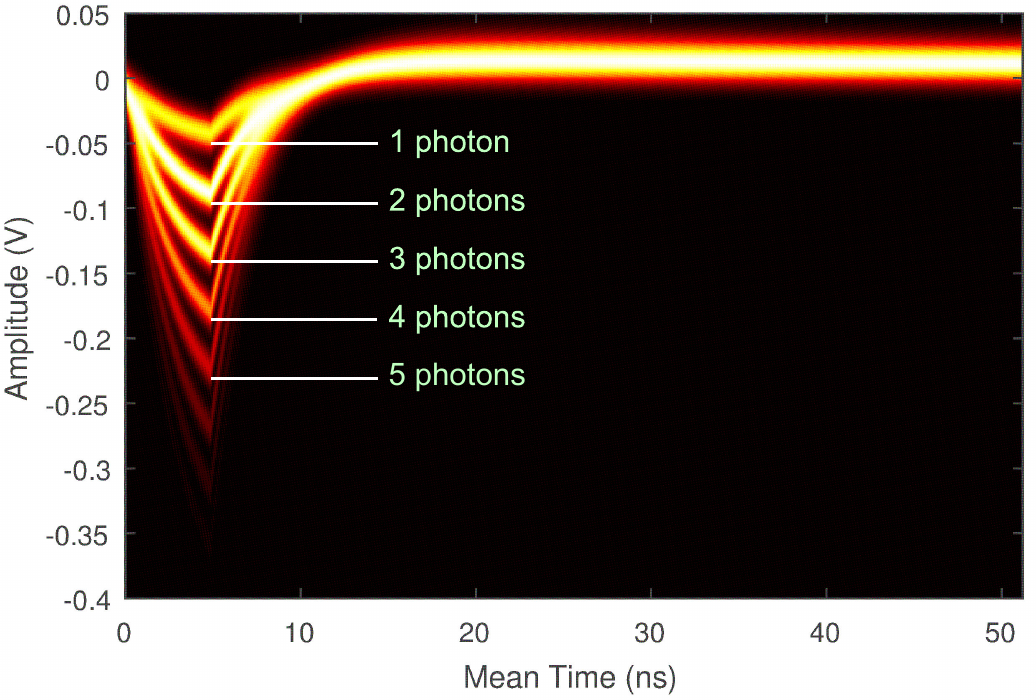}%
\caption{Responding intensity graph integrated by pulse shapes.Each photodiode element in the array is referred to as a "microcell". Typically numbering between 100 and 1000 per mm$^2$, each microcell has its own quenching resistor. The signals of all microcells are then summed to form the output of the SiPM.}
\label{Imhomo}
\end{figure}

\begin{figure}[!ht]
\center
\includegraphics[width=3 in]{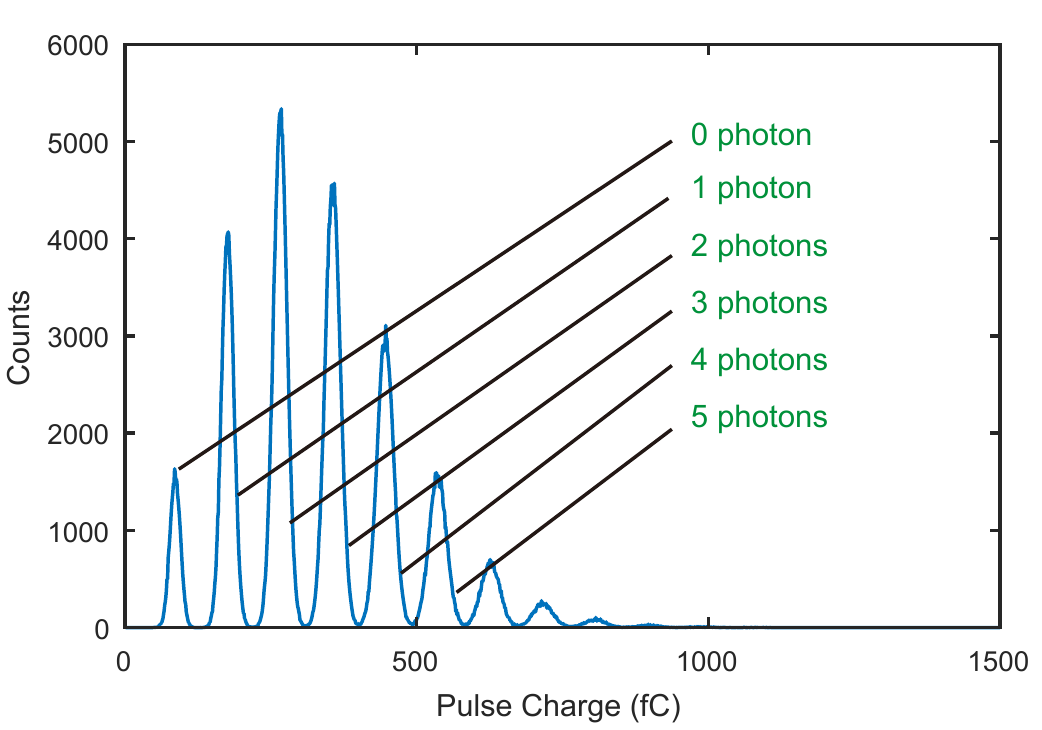}%
\caption{The single photons' energy spectrum.To overcome proportionality lack in APD, the SiPM integrates a dense array of small, electrically and optically isolated Geiger-mode photodiodes.Each microcell detects photons identically and independently. The sum of the discharge currents from each of these individual binary detectors combines to form a summed photons output, and is thus capable of giving information on the magnitude of an incident photon flux.}
\label{EnergySpectrum}
\end{figure}

To overcome proportionality lack in APD, the SiPM integrates a dense array of small, electrically and optically isolated Geiger-mode photodiodes. Each photodiode element in the array is referred as a "microcell". Typically numbering between 100 and 1000 per mm$^2$, each microcell has its own quenching resistor. The signals of all microcells are then summed to form the output of the SiPM. Each microcell detects photons identically and independently. The sum of the discharge currents from each of these individual binary detectors combines to form a summed photons output and is thus capable of giving information on the magnitude of an incident photon flux. A spectrum of the same pulse is shown in Fig. \ref{EnergySpectrum} , and the response to low-level light pulses is shown in Fig. \ref{Imhomo}.

\subsection{Pulse shape and SiPM Band Width}
\begin{figure}[!ht]
\center
\includegraphics[width=3.4in]{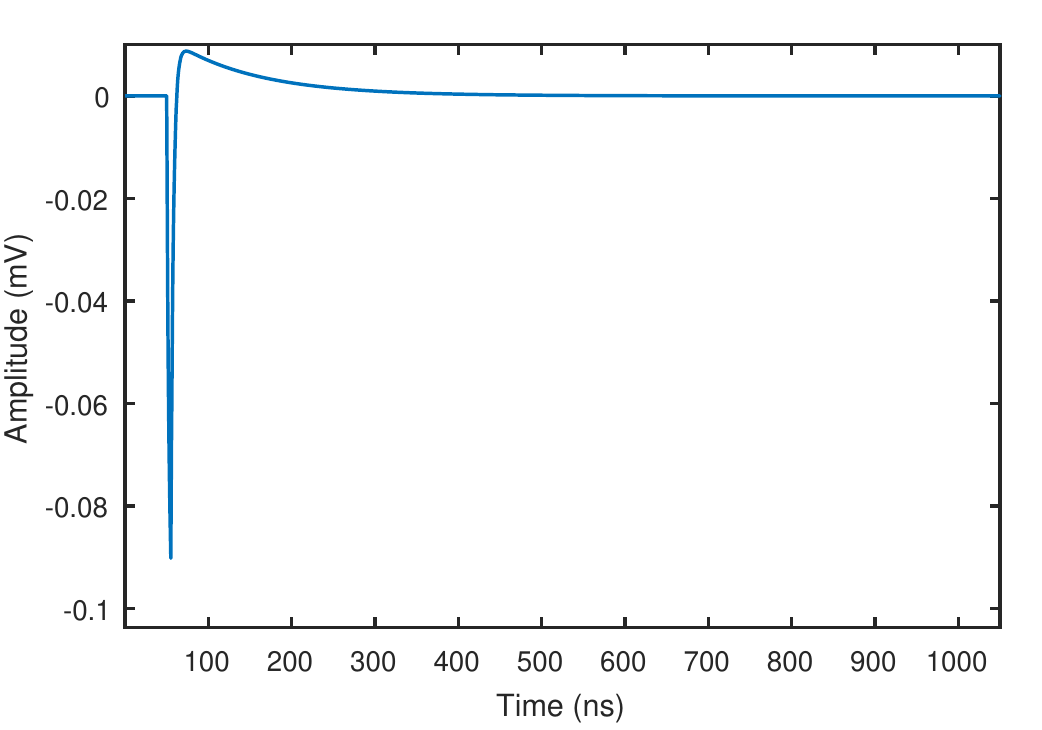}%
\caption{ Mean pulse of the SiPM pulse. Here, the mean pulse is obtained from averaged pulses output by the SiPM on low event rate. Low event rate pulses avoid pileup case. The mean pulse has a sharp pulse peak with fast leading and recovery edge.}
\label{meanPulse}
\end{figure}

\begin{figure}[!ht]
\center
\includegraphics[width=3.4in]{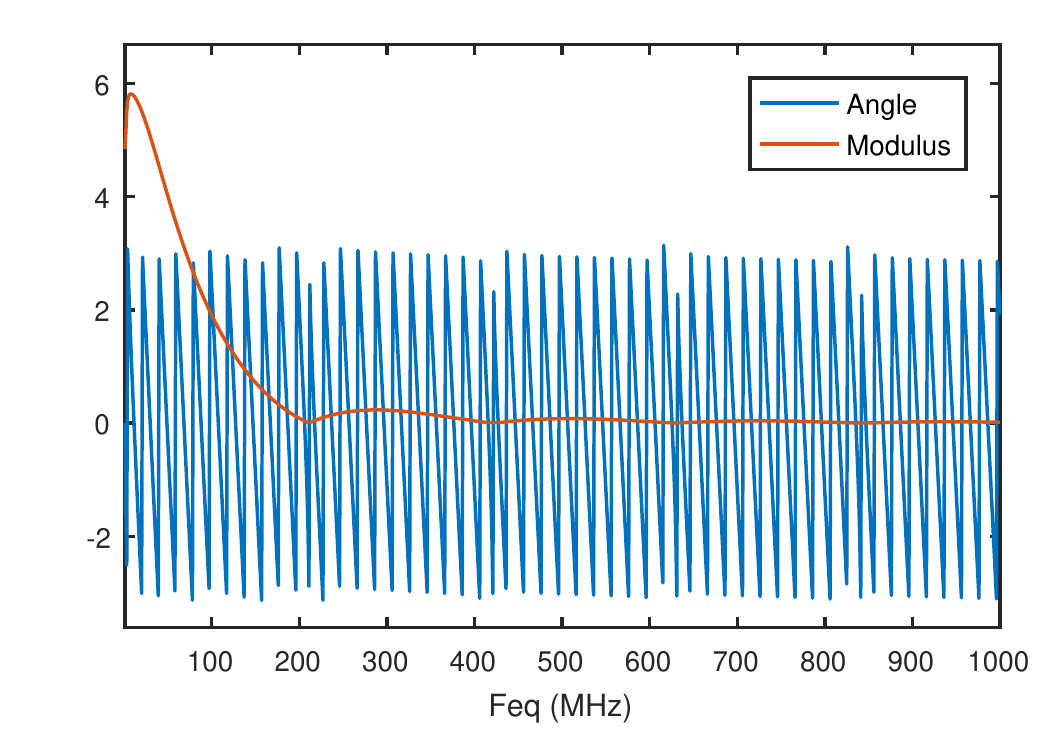}%
\caption{ Frequency property of the SiPM pulse. The VLC system transfer data using the repeating pulses amplitude. So the frequency width is expected high for more transfer data rate.}
\label{meanPulseFFT}
\end{figure}

In the VLC system, different optical pulses, are fed to SiPM, and generated by LED, and then produced the pulses chain. Transferred data are modulated at the height of optical pulses. We find the allowing data rate without errors depends largely on the single pulse response of the SiPM output. Sharper the SiPM response, more possible transfer rate can be afforded. To know the SiPM response bandwidth, we evaluate the SiPM pulse and its frequency response. Fig. \ref{meanPulse} shows the impulse response of SiPM, which have a sharp pulse peak with fast leading and recovery edge. And we employ Fast Fourier Transform to obtain its frequency response in Fig. \ref{meanPulseFFT}. In Fig. \ref{meanPulseFFT}, we also can see the cut-off frequency that exceeds 100 MHz in Fig. \ref{meanPulseFFT}, which is significantly better than common photon devices.

\subsection{Reconstruction Precision and Required Time}
It also can be noted that recovery edge is followed by positive tails in the SiPM pulses. Then pulse aliasing cannot be avoided when the repetition rate is enough to keep the data transfer rate. Thus we employ three different reconstruction methods to inverse the data input: SR, PMID, and FD. When carrying out the calculation, 100 iterations PMID elapsed more than $10^5$ times which takes more computing time than SR method. And all the 1000 MByte pulse chain with 10 Gsps sample rates is too high to encode by PMID method. To obtain the reconstruction precision of the three methods, we calculate the residual error for just 100 MByte pulse chains.

\begin{figure*}[!ht]
\center
\includegraphics[width=5in]{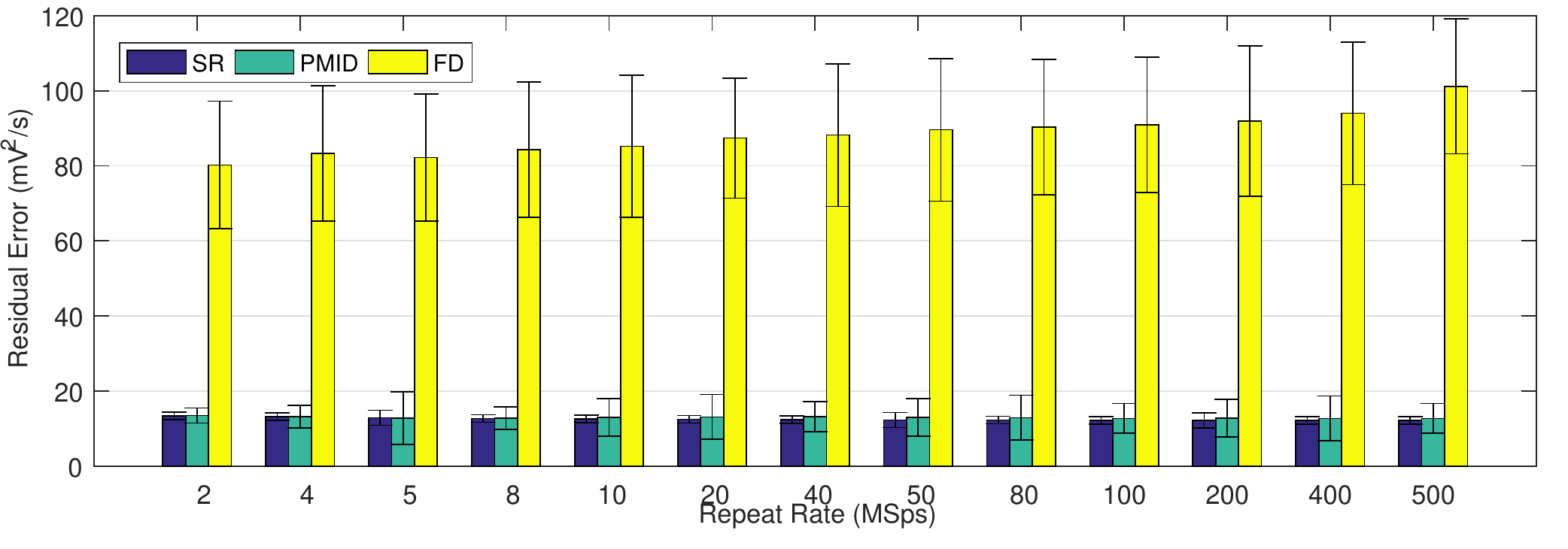}%
\caption{ The residual errors produced by SR, PMID and FD method.Residual Error here means the difference between received sample and theoretical value calculated from reconstruct data ,is calculated by Eq. \ref{ResidualErrorComp}.}
\label{ResidualError}
\end{figure*}

\begin{figure}[!ht]
\center
\includegraphics[width=3.5in]{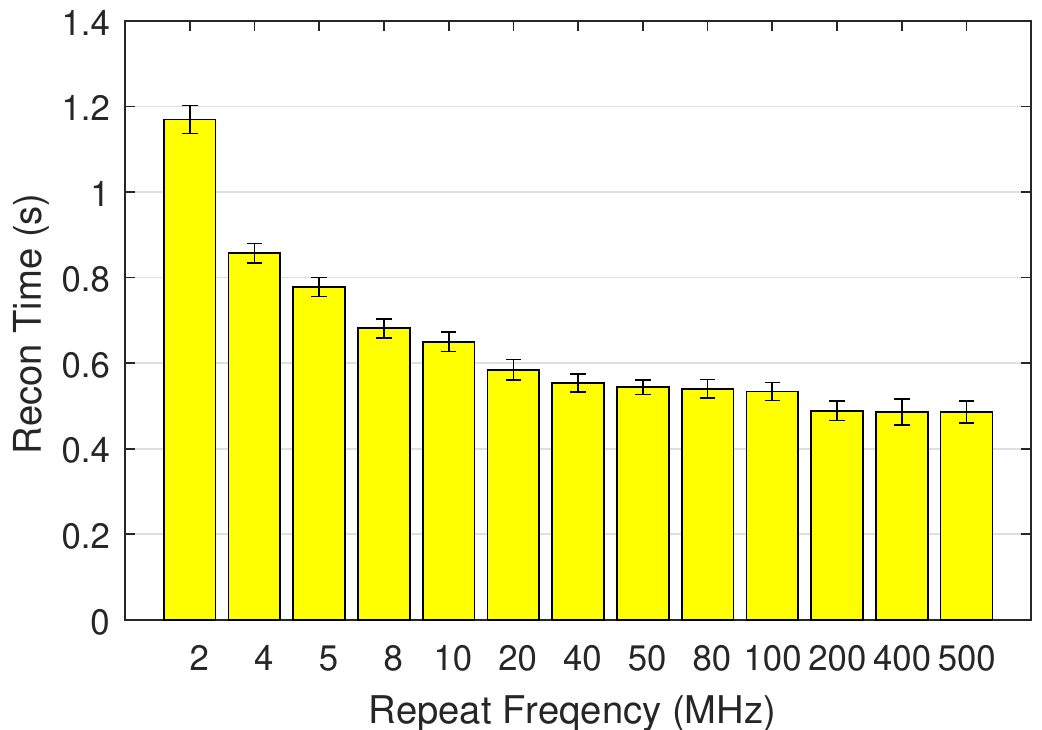}%
\caption{ Required time to restore the pulses chain containing 1 Mbytes using SR method. These time values are calculated on computer workstation.  The comparison of elapsed time was based on the hardware of Intel Core i3-3220@3.30GHz and 8GB 1600MHz Kingston DDR3 memory. The program was implemented in MATLAB. }
\label{plotrecontime}
\end{figure}

In Fig. \ref{ResidualError}, Residual Error of each method is evaluated. Residual Error here means the difference between the received sample and theoretical value calculated from reconstruct data, which is calculated by Eq. \ref{ResidualErrorComp}.

\begin{equation}
\label{ResidualErrorComp}
r = |\overrightarrow{p} - H\overrightarrow{f^{\ast}}|^2,
\end{equation}
where $f^{\ast}$ is the vector containing reconstruct data. With the increase of Repeat Rate, Residual Error of three algorithms keeps rising. By contrast, SR and PMID algorithm tend to be stable, and is not affected by Repeat Rate too much. We can see that SR and PMID produce one order of magnitude lower residual error than FD. Considering SR processed data free of whole data iteration, we believe the SR is the appropriate method in the practical system. Since each PMID iteration needs 2 convolution operators and 3 vector multiplication operators, it has large computing burden when the iteration number and sample rate increased. We also find the required reconstruction time is affected by the pulse repetition rate. In Fig. \ref{plotrecontime}, we investigate the reconstruction time of SR method. We find the required reconstruction time is decreased as the repetition rate increases. This can be illustrated as that fewer samples need processing when the repetition rate increases.

\subsection{Required Sample Rate}
\begin{figure}[!ht]
\center
\includegraphics[width=3.5in]{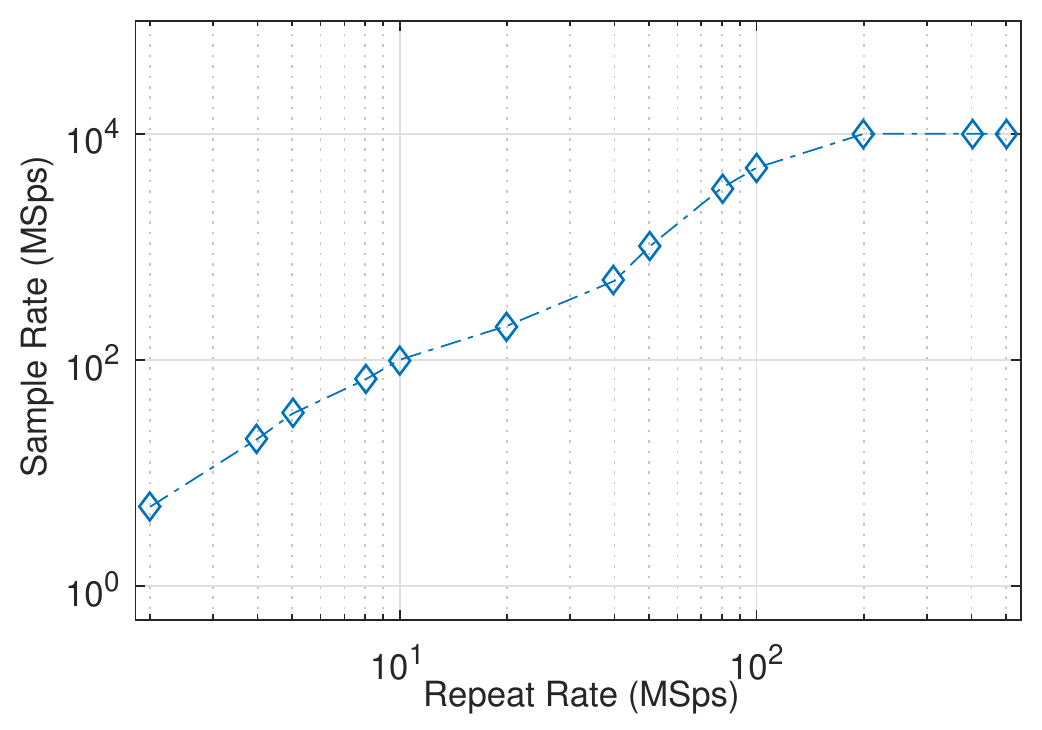}%
\caption{Required minimal sample rate of digitizer to guarantee 1000 Mbytes all right restored pulses amplitude. We investigated the sampling rate's effect on the SiPM VLC system. Data points were picked up at different intervals from the original sequence to simulate different Analog-to-Digital Converter(ADC) sampling rates, ranging from 2 Mega-Samples per second (MSps) to 10 GSps.}
\label{minSampleRate}
\end{figure}

The required sample rate of SiPM digitizer is another critical aspect in VLC systems. When the sample rate of the digitizer is insufficient, the digitized signal sample cannot guarantee the correctness of pulse reconstruction. And when the sample rate of the digitizer is too high, the cost and the computation time will be both unaffordable. The above results are obtained from data sampled at 10 GSps. The pre-stored pulses data with a high sampling rate allows the electronic designer to evaluate their DSP algorithms with good operability and flexibility. However, 10 Giga-Samples per second (GSps) is not necessary for the VLC system to process the pileups. We investigate the sampling rate's effect on the SiPM VLC system. Data points are picked up at different intervals from the original sequence to simulate different Analog-to-Digital Converter(ADC) sampling rates, ranging from 2 Mega-Samples per second (MSps) to 10 GSps. Thereafter, the re-sampled waveform is screened and processed by the SR method. We investigate the minimum required sample rate of digitizer which all guarantees 1000 Mbytes pulses amplitude totally and correctly restores using re-sample data experiment as shown in Fig. \ref{minSampleRate}. In Fig. \ref{minSampleRate}, when Repeat Rate is 10MSps, the sampling rate can reach 100MSps, and considering the point Repeat Rate 100MSps, the sampling rate value of this point is about $10^4$ MSps, which proves that our system can achieve high data rate when sample rate is sufficient. Overall, the required sample rate increases with the pulse repetition rate. This result suggests that the sample rate of SiPM digitizer can be vital when the superior transfer rate is expected.

\section{Conclusion}
Based on the experience of SiPM application in the imaging system, we constructed a SiPM based VLC system and thus transferred the data transmission problem to a signal processing problem. Thereafter, we applied the SR algorithm to solve this problem. Experiments with real signals showed that this method provided good signal pulse fitting at a different repetition rates. The pulse reconstruction performance of this method was also better than the previously used PMID and FD methods.

The proposed and implemented system has 4 Gbit/s transfer rates under one 10 GSps DA/AD channels. This system does not require the mean pulse of signals to be in a particular shape, as long as it is fixed. Since different front-end electronics in detection systems may produce different pulse shapes, the adaptability of the proposed method is appropriate in applications. We preliminarily conclude the high transfer rate of SiPM based VLC system attributes the high bandwidth of SiPM signal pulse and single photon counting ability.

In the future work, we will apply the SiPM to different color visible light and multiple voltage threshold digitizers, instead of a single color LED and the high-speed ADC. Since both Multi-Voltage threshold method and the SR method require the prior knowledge of pulse shape, they could be well-matched. Other iterative algorithms, such as maximum a posterior (MAP) or ordered subset expectation maximization (OSEM), can also be considered in the pulse recovery process.

\begin{appendices}

\begin{table}
\caption{Abbreviations table}
\label{Abbreviation}
\centering
\begin{tabular}{c c}
\hline
Abbreviations &  Full Name \\
\hline
VLC & visible light communication \\
PD & photo diode\\
LED&light-emitting diode \\
SiPM&Silicon photomultiplier\\
PMT&Photomultiplier tube\\
PDE&Photon Detection Efficiency\\
APD&Avalanche Photodiode\\
PIN&Positive Intrinsic-Negative\\
PET&positron emission tomography\\
PMID&pulse model based iterative deconvolution\\
FD&Fourier deconvolution\\
SR&Successive Remove\\
GSps&Giga-Samples per second\\
MSps&Mega-Samples per second\\
ADC&Analog-to-Digital Converter\\
\hline
\end{tabular}
\end{table}
\end{appendices}

\bibliographystyle{IEEEtran}
\bibliography{bibSiPMVLC}

\end{document}